# Knowledge-generating Efficiency in Innovation Systems:

## The relation between structural and temporal effects

Inga Ivanova [a] and Loet Leydesdorff [b]

[a *] corresponding author; School of Business and Public Administration, Far Eastern Federal University, 8, Sukhanova St., Vladivostok 690990, Russia; inga.iva@mail.ru

[b] Amsterdam School of Communication Research (ASCoR), University of Amsterdam
PO Box 15793, 1001 NG Amsterdam, The Netherlands; loet@leydesdorff.net

**Abstract**

Using time series of US patents per million inhabitants, knowledge-generating cycles can be distinguished. These cycles partly coincide with Kondratieff long waves. The changes in the slopes between them indicate discontinuities in the knowledge-generating paradigms. The knowledge-generating paradigms can be modeled in terms of interacting dimensions (for example, in university-industry-government relations) that set limits to the maximal efficiency of innovation systems. The maximum values of the parameters in the model are of the same order as the regression coefficients of the empirical waves. The mechanism of the increase in the dimensionality is specified as self-organization which leads to the breaking of existing relations into the more diversified structure of a fractal-like network. This breaking can be modeled in analogy to 2D and 3D (Koch) snowflakes.

The boost of knowledge generation leads to newly emerging technologies that can be expected to be more diversified and show shorter life cycles than before. Time spans of the knowledge-generating cycles can also be analyzed in terms of Fibonacci numbers. This perspective allows for forecasting expected dates of future possible paradigm changes. In terms of policy implications, this suggests a shift in focus from the manufacturing technologies to developing new organizational technologies and formats of human interactions

## 1. Introduction

The explanation of economic changes in terms of underlying mechanisms has been focal to evolutionary economics [1, 2]. According to Schumpeter [3] the development of the economy is based on continuous innovations. This is especially true for the post-industrial stage where the proliferation of knowledge is considered as an important source of consistent growth [4, 5]. When studying Japan, Freeman [6] first noted, that knowledge-generating can only be economically successful if an innovation system is in place. Lundvall [7, 8] and Nelson [9] elaborated on a systems perspective in innovation studies. Porter [10, 11] abstracted from the national context by focusing on "clusters" of innovations which can be differently shaped in regional and/or national settings. Gibbons *et al.* [12] distinguished between a knowledge-production paradigm in niches such as universities ("Mode 1") and trans-national and trans-disciplinary knowledge production ("Mode 2"), that is driven by communication across institutional borders. "Mode 2" was further elaborated in terms of University-Industry-Government collaborations as the Triple Helix model [13, 14].

Initially, the concept of an innovation system was developed with a focus on national system of innovations. In later studies, one introduced the notion of smaller-sized innovation systems, such as regional [15, 16], sectoral [17, 18], technological innovation systems [19, 20], and corporate innovation systems at different scales [21]. A national system of innovations, as in the case of Hungary, can be comprised of a number of smaller regional systems [22]. This concept of nested innovation systems was also proposed as a model for economic development at the city level [23].

The systems perspective relates to evolutionary theorizing because a system is shaped when different selection mechanisms can operate upon one another. Two selection environments can mutually shape each other in a coevolution along a trajectory, but adding a third sub-dynamic can cause a bifurcation and consequential transition in the system at the regime level [24]. Ivanova & Leydesdorff [25] argued that adding a third selection environment to a system established in terms of bi-lateral (e.g., university-industry) relations or bilateral (e.g., government-university) policies can drastically change the behavior of an innovation system, because a third sub-dynamics provides an additional source of variation that continuously upsets previous tendencies towards equilibrium [1]. Note that this accords with Simmel's observation that the difference between social dyads and social triads is fundamental and this difference refers not so much to the number of participants as to more fundamental issues, such as quality, dynamics, and stability of the resulting system [26].

When referring to the innovation activity of a system, one can use the notions of the *capacity* and *efficiency* of an innovation system in order to differentiate among different systems and inform policy choices. Various quantitative methods in which a number of input indicators are used to calculate output indicators have been developed to evaluate the *capacity* of innovation system. These methods can be categorized into three categories: Composite (Innovation) Indicators, Data Envelopment Analysis (DEA), and Modeling/Econometric Approach [27].The *efficiency* of an innovation system, however, is difficult to specify, because of the complexity of and possible synergy among innovation activities, such as investments in R&D, the numbers of new services and products, patents, researches, etc. [28]; adequate efficiency indicators are difficult to construct. Another reason is that innovation statistics is still rather uncertain which leads to much stochastic fluctuations and consequently difficulties in the parameter estimation.

Patents have been used as a simplification of innovation indicators. But there is no one-to-one correspondence between patents and innovations. Only a small percentage of patents can be expected to be used in practice, and only a small percentage of patents used can be expected to pass to the category of innovations. The drawback of using patent indicators is that they are very uncertain in representing innovation output [29]. However, patents can be used as a measure of the intensity of innovation activity [30].

The efficiency of an economic system can be defined analogously to technical efficiency as the ratio of output to input [31]. An innovation system can be considered as efficient if it is able to produce the maximum possible output from a given amount of innovative input. Efficiency can then be defined using the knowledge production function (KPF) with the number of patents as an output variable [32, 33] that can be written as a product of input variables [34, 35, 36]. The input variables can be rather diversified, such as the level of R&D expenses, the number of R&D employees, the state of the technological, industrial, and institutional infrastructures, etc. However, one cannot encompass all the factors that influence the capacity of an innovation system, because some of these factors cannot be measured. For example, the interaction between the different elements of an innovation system may generate self-enforcing (auto-catalytic) systemic effects that affect the performance of the system [37].

When comparing innovation systems at national or regional levels, it can turn out that two systems perform unevenly despite a set of equal input parameters. One would then theoretically expect a more equal efficiency. This discrepancy can be attributed to differences in the intensity and quality of interactions in the systems under study. In other words, one risks comparing non-comparable systems such as systems of a different nature or with different structural organizations. The mechanism of self-organization, lying at the origin of biological complexity, can be also expected to provide system change in the economy [38] by generating more sophisticatedly organized and efficient systems under selection pressure. In summary, one can expect a relation between the organizational efficiency of a system and the level of the system's self-organization.

Our research question is to explore the influence of complexity in innovation systems on the knowledge-generating performance, and regularities in the improvements of the efficiency of the system over time. To that end we compare measurement results with maximal efficiencies that can be derived from the theoretical model. The analysis is pursued at the macro level of the system. The paper is organized as follows. In Section 2 statistical patent data (USPTO) are analyzed; four distinctly shaped cycles are distinguished for the years 1840-2013. These cycles partly coincide with Kondratieff cycles. A model of efficiency of knowledge generation in innovation systems is developed in Section 3. The model explains the empirical findings in considerable detail. A conclusion of this model is that the system's performance is proportionate to the complexity of the system. In Section 4, we outline the perspectives of the model extension to a next-higher dimensionality in order to specify expectations. In Section 5, the results are summarized and options for policy-makers are elaborated in Section 6. The mathematical derivations for calculating the dimensionality of innovation systems are provided in two Appendices.

## 2. The data

The newly generated technologies can be considered as inputs to the total productivity together with other knowledge carriers [39, 40]. The number of patents can be considered as an indicator of the innovation capacity of a system [41]. Patents reflect the innovations in databases for the purpose of legal protection of intellectually property. Furthermore, there is a correlation between innovation capacity and country's overall competitiveness and level of prosperity [30]. The economic growth implies the growth of innovation efficiency, and the increase in the number of patents can be made visible by patent statistics.

Korotaev *et al*. [42] suggested that the dynamics of the number of patents granted annually in the world per million inhabitants would show the patterns of Kondratieff cycles for the period 1900-2008. However, there is no single and unique periodization of the long-wave Kondratieff cycles available. One possible periodization is as follows: the first cycle spans from 1780 till 1848; the $2^{nd}$ cycle runs from 1848 till 1895; the $3^{rd}$ cycle lasts from 1895 till 1940; the $4^{th}$ cycle from 1941 till 1973; and the $5^{th}$ cycle starts from 1973 [43]. Economic cycles are reported to have strong relationship with technological innovations, and there can be significant differences in economic cycles between two countries [44].

For example, the distinctions are far from obvious in US patents dynamics for the period 1840-2013 shown in Figure 1. One can no longer discern the second and third waves; they are merged in a single wave, and one can perhaps distinguish an additional sub-wave for the period 1980-2006. The patent statistics for period 1840-2013 were retrieved from United States Patent and Trademark office (USPTO),[1] and data on US population dynamics were taken got from the website at http://www.populstat.info/Americas/usac.htm.[2]

---

[1] http://www.uspto.gov/web/offices/ac/ido/oeip/taf/h_counts.htm
[2] Accessed on April 7, 2014.

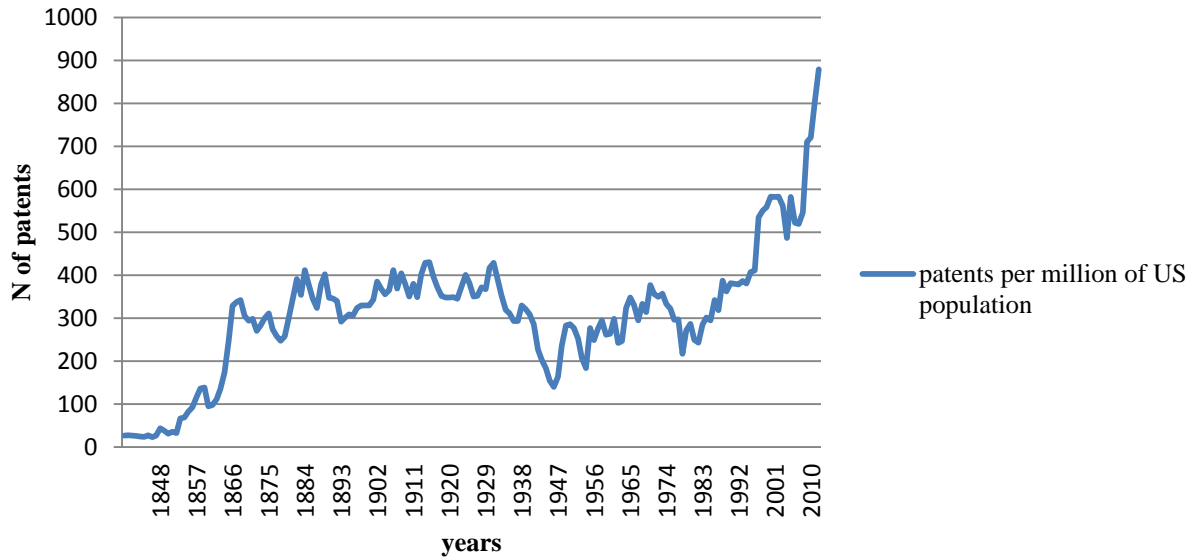

**Fig. 1**: Dynamics of the numbers of US-patents granted per million inhabitants of the USA during the period 1840-2012.

The mismatch between Kondratieff waves and patent waves can be attributed to differences among prevailing knowledge-generating paradigms. From the picture presented in Figure 1 one can distinguish four periods: 1840 - 1945; 1945 - 1980; 1980 - 2006; and from 2006 onwards. Let us assume that the mechanisms for knowledge generation in these four periods are different. Figures 2 – 4 separately show the first three of these four periods. Average efficiency of knowledge generation can then be evaluated by the number of patents generated over the respective time period; this can be mapped by the slope of a linear regression, using Ordinary Least Square (OLS), and indicated by a straight line. For the fourth period (Figure 5) which begins in the year 2006, there is not sufficient data, so that one can measure the efficiency of the corresponding system only approximately.

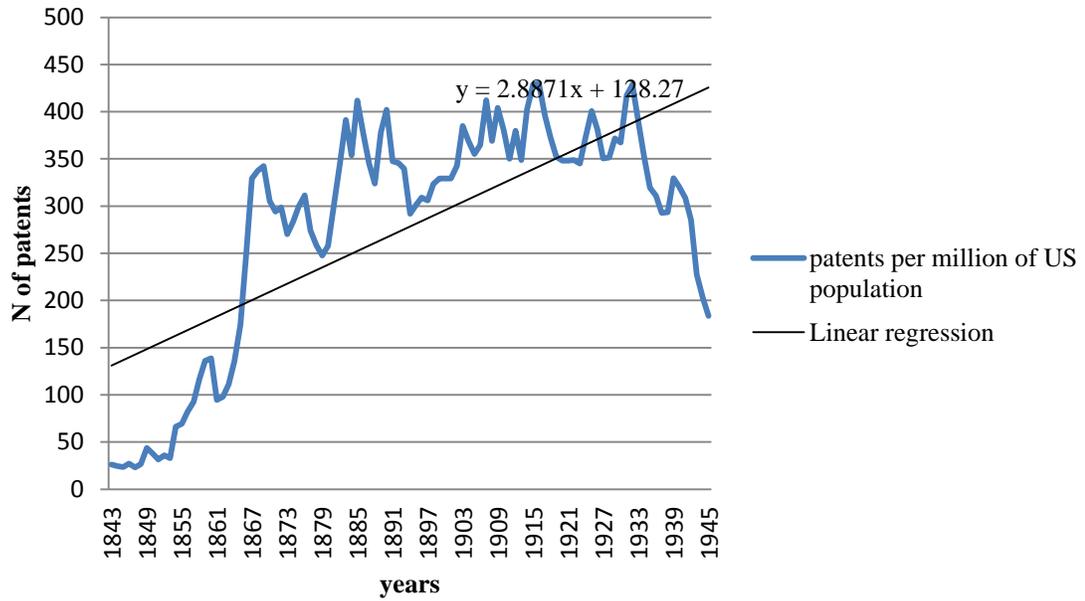

**Fig. 2**: Dynamics of the numbers of US-patent granted per year per million inhabitants of the USA during the period 1840-1945; the straight line is based on linear regression.

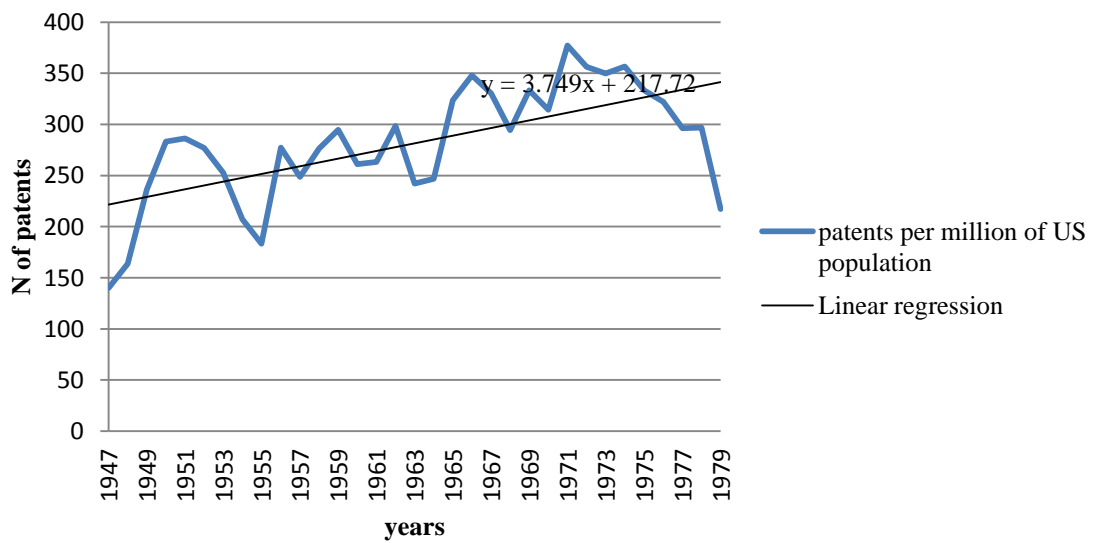

**Fig. 3**: Dynamics of the numbers of US-patents granted per million inhabitants of the USA during the period 1945-1980; the straight line is based on linear regression.

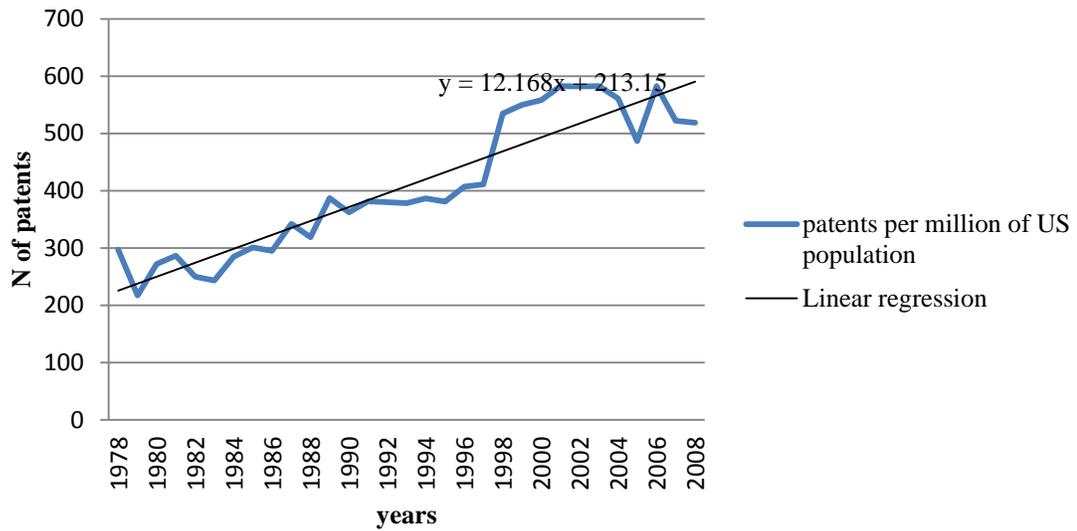

**Fig. 4:** Dynamics of the numbers of US patents granted per million inhabitants of the USA during the period 1979-2006; the straight line is based on linear regression.

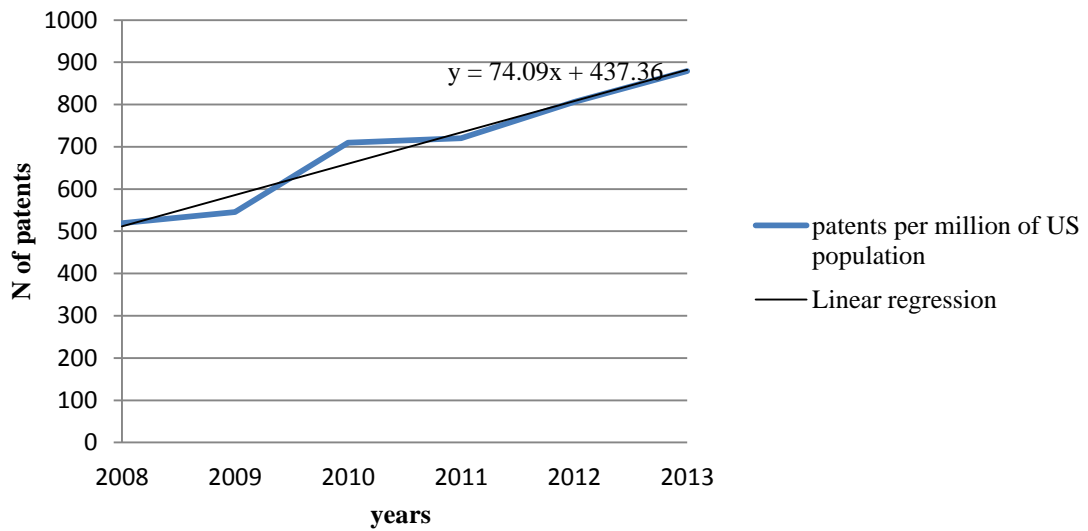

**Fig. 5**: Dynamics of the numbers of US patents granted per million inhabitants of the USA during the period from 2006; the straight line is based on linear regression.

The efficiency of knowledge generation for the periods in question successively increases by the factors: $k_2/k_1 = 3.749/2.887 = 1.3$; $k_3/k_2 = 12.168/3.749 = 3.24$; $k_4/k_3 = 74.09/12.168 = 6.09$. We note that the period 1840-1945, mapped in Figure 2, covers the second and the third Kondratieff cycles. Figures 3 and 4 refer to the fourth and a part of the fifth cycle, respectively. The change of innovation paradigms, displayed in Figures 2 and 3 can be attributed to the shift in innovation sources. During the period before WW II (Figure 2) these sources were mainly confined to single institutional spheres. After WW II, innovations were likely to emerge from various mechanisms of bi-lateral inter-institutional interactions, such as university - industry cooperations.

The fifth cycle (starting approximately in 1980) begins at about the date when the focus of US science, technology, and innovation policies shifted from heavy industry to a new technology- and knowledge-based economy [45]. The growth rate of knowledge-based industries in the years1986-1996 in the EU and the US exceeded the average rate of growth of the business sector, and the share of these sectors in business added value reached 50% by the mid-90s [46]. This was in response to the phase of economic globalization that began during 1980s [47]. In the US the average annual increase of knowledge-workers amounted to 3.3 % for the period 1992 - 1999, compared to 2 - 3% for the previous decade [48].

The knowledge-generating paradigm also changes during this period; for example, from university – industry dyads to triadic relationships and other forms of trans-institutional collaborations [49, 14, 12]. Although this paradigm shift is global, the empirical transitions in individual countries can be different. Time lags between the leading and catching-up countries are inevitably relative to the needed policy adjustments and profound changes in national innovation systems [50].

**3. The model**

Let us explore additional drivers affecting the knowledge-generating capacity of innovation systems. Our argument will be that the capacity of innovation system is proportional to the fluxes of exchange which operate *in* network links (that is, the links as dyadic channels for the exchange) and *on* network links (that is, the links are subject to change as a result of triadic or higher-order interactions). The interactions among systemic selection environments can provide a synergy [51].

An innovation system, comprising more than a single actor, can be considered as a dynamic system, where innovative dynamics can be expected to change over time due to the change of the intensity and quality of the interactions. We measured above (in Figures 2-5) the average efficiency for the respective time periods. The potential efficiency of a system, however, depends on all possible states that a system can take. This can be defined as the sum of possible states of the system combined with corresponding weight factors. All possible states of the system can be represented as a phase space, with each possible state corresponding to a point in this space. The dimensionality of the phase space is defined by the system in question and can vary.

At each point of the phase space *s* the system exhibits a different synergy and a different performance. This can be accounted for by a weight function *f(s)*. The theoretical performance of the system can then be specified as the sum of different possible synergies in accordance with Eq. 1, that is, by taking the integral over the phase space as follows:

$$P = \int f(s)ds = \bar{f} \cdot \int ds \tag{1}$$

In Eq. 1, $\bar{f}$ is the average value of the weight function. The phase space can be represented by a line, surface, volume, etc., depending on the dimensionality of the system.

For the sake of simplicity, let us consider first the case of an innovation system of bilateral university-industry relations. Abstractly, this double helix (DH) can be drawn as in Figure 6, in the form of two intercepting institutional spheres.

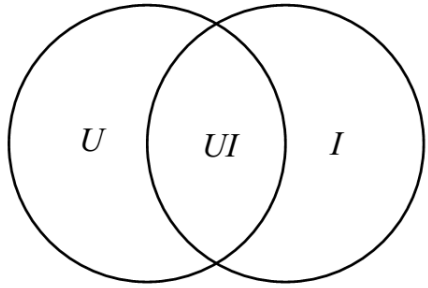

**Fig.6:** A model of dyadic university (*U*) - industry (*I*) relations

The overlapping area corresponds to the sphere of university-industry interactions where the actors partially can replace each other in their respective functions of wealth generation and novelty production. This model can alternatively be mapped into a Cartesian coordinate space (Figure 7). In this representation, the unit vector *V* has coordinates along the two axes *U* and *I*. The vector accounts for the relative roles played by the two institutional spheres.

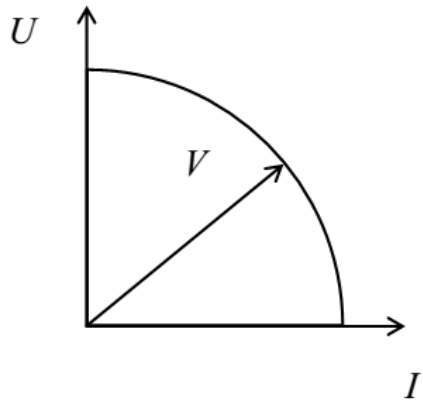

**Fig.7:** Cartesian coordinate representation of a "double-helix" model of university-industry interactions.

The relative roles may change over time. The arc drawn by the end of vector V represents all the possible contributions of actors or the phase space of this system. According to Eq. (1) one obtains $P_2 = \bar{f}_2 \cdot \pi/2$. For a system with a single institutional sphere the coresponding capacity would be $P_1 = \bar{f}_1$. Assuming that $\bar{f}_1 = \bar{f}_2$, the systems relative efficiency can be calculated as the ratio $P_2/P_1 = 1.57$. This value differs only 17% from with the ratio $k_2/k_1 = 1.3$, obtained as a quotient of the regression coefficients in Figures 2 and 3 (above).

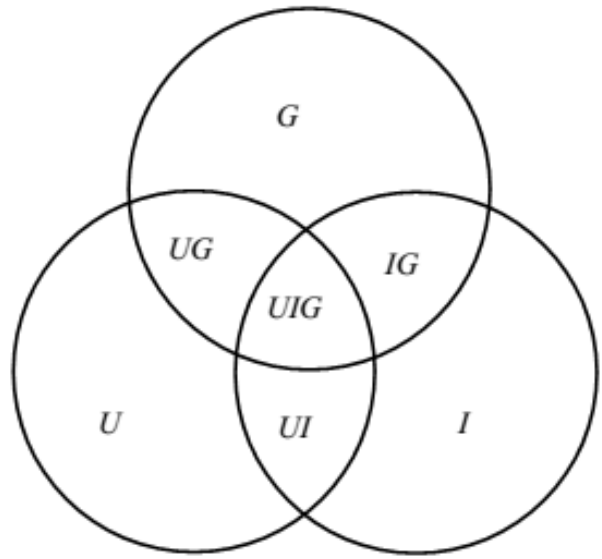

**Fig.8**: Representation of a Triple Helix model of University (*U*), Industry (*I*), Government (*G*) relations

The above analysis can be elaborated to network arrangements in an innovation system formed by three selection environments operating or a Triple Helix (TH) model. Figure 8 shows this model using a representation equivalent to the one in Figure 6: the circles marked by capital letters *U*(niversity), *I*(ndustry), and *G*(overnment) represent the corresponding institutional spheres responsible for the three functions of novelty production, wealth generation, and normative control, and their overlapping areas of dual and triple interactions [52].

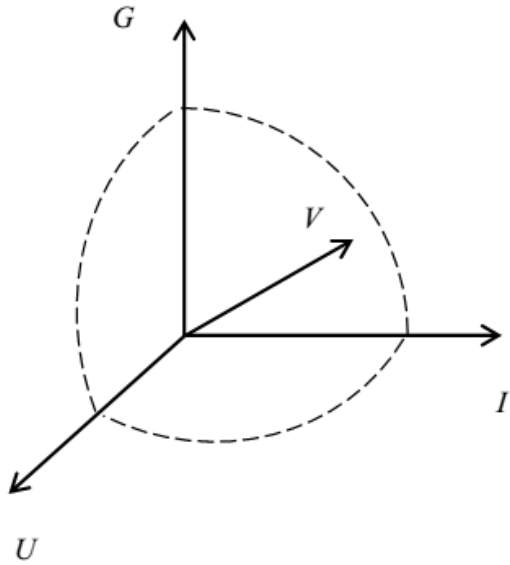

**Fig.9:** Cartesian coordinate representation of the TH model

The TH system can similarly be presented in a Cartesian space as a vector *V*, drawn from the origin of the coordinate system (Figure 9; [25]). Projection of the vector *V* on the axes *U*, *I*, and *G*, corresponds again to the relative roles played by the different institutional spheres. Each agent's relative role can be defined in terms of relative partaking in novelty production, wealth generation, and normative control insofar as generated by this actor. Over time, however, the relative contributions of the institutional spheres can be expected to change. This can be accounted for by a rotation of vector *V* in the coordinate space. Such a rotation in this case reflects the variations of the self-organizing dynamic of interactions among the three selection environments.

The TH phase space, which represents all possible states of three-lateral interactions, is denoted by dashed lines in Figure 9, and corresponds mathematically to a 2D-spherical surface, limited by a three planes: *UG*, *GI*, and *UI*. Projected onto the plane, this surface resembles a convex triangle. For the sake of simplicity, this convex surface triangle is drawn as an equilateral triangle in Figure 10.

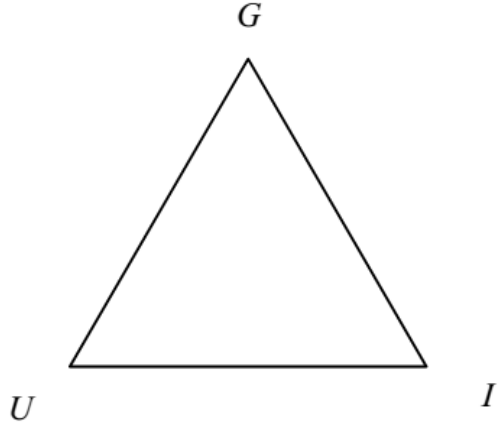

**Fig.10:** Simplified representation of the TH phase space

The capacity of a TH system can be calculated according to Eq. (1), by integrating over the spherical surface. Provided that the surface area of the TH phase space equals $\pi/2$, one obtains $P_3 = \bar{f_3} \cdot \pi/2$. If one assumes that $\bar{f_3} = \bar{f_2}$ then the capacities of DH and TH system would be the same, but this contradicts the results obtained from the statistical data above.

We suggest that this obvious discrepancy of the results can be explained by taking the non-linear dynamics of TH behavior into account. In the case of three dynamics, the operation is "non-Abelian" [25]: the order in the sequences of steps or, in other words, the direction of the rotation matter for the outcome. One can also say that this system is path-dependent. From a mathematical perspective, a TH can be expected to self-organize at the network level by creating clones of itself at different scale levels and in all three directions. This fractal propagation is intrinsic to networks of three-lateral and higher-dimensional relations.

In addition to relations, correlations are now important, and correlations can be spurious. This makes the development non-linear. Consequently, the TH network can be expected to break into a fractal-like structure when proliferating. Therefore, a TH system can be more densely "nested" in a knowledge based economy than a relational (DH) system. Due to the fractal nature of the complexity, a TH network can be expected to exceed the complexity of DH network by an order of magnitude. This next-order non-linearity can also be expected given quadruple or higher-order selection environments.

Using the TH metaphor, the procedure of this self-replication can be depicted as follows. The activity spheres—that is, selection environments—of the key actors can increasingly overlap. Universities, alongside with their intrinsic educational and research functions, can then undertake business functions, such as creating small innovative companies. Industrial corporations can create their own research centers and training centers for employees. They may also use the university's infrastructure for elaborating their R&D projects. In a similar manner corporations may shift part of their R&D projects costs to the state, since the state remains the main source of funding for universities. Governments in turn encourage the development of small innovative enterprises through priority financing of specific universities and legislative regulation. They can also incentivize industry to develop and implement new innovative technologies.

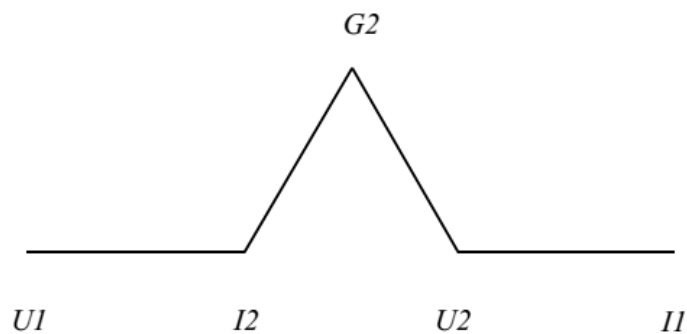

**Fig.11:** Second-order self-similar pattern of TH model

The relations between universities and industry can partially substitute for the state in the creation of an innovation infrastructure (e.g., [53]).  Not only the corners of the triangle, but also the sides—representing interactions—can thus be (second-order) interactive. For instance, university *U1* initially engaged in arrangements with industrial corporation *I1*, can create one or more spin-off firms. These daughter firms can be expected to shape themselves in compliance with the form of their parental structures, possessing second-order novelty production *U2*, wealth generation *I2*, and normative control *G2* units, but in a different mix. Graphically the multitude of second-order spin-off firms can be depicted as a smaller-size triangle formed on a side of the larger triangle (Figure 11).

This process can be iteratively continued forming a TH network as a manifold. This proliferation can be compared to the construction of a fractal structure known as the Koch snowflake which is the limit of an iterative process that starts with an equilateral tiangle and comprises infinite dividing of each line segment into three segments and drawing equilateral triangle on the base of the middle segment [54]. Four first pre-fractals of the Koch snowflake are shown in Figure 12.

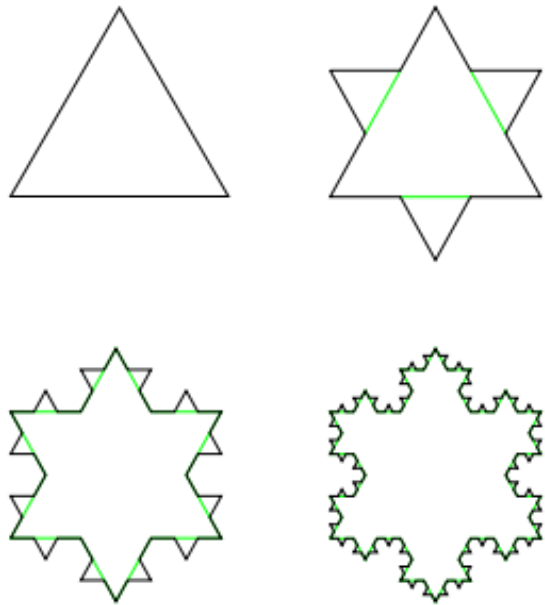

**Fig.12**: Four first pre-fractals of the 2D Koch snowflake

Due to the non-linear interactions within a TH, the surface area of the phase space increases endogenously. This accordingly can lead to an increase in the system's maximum efficiency as follows: The surface area of each next triangle in the iteration, or equivalently the average efficiency of the next level innovation system, equals approximately 15.8% of the previous one, and the TH phase space surface area exceeds the surface area of the original triangle by a factor 3.3 (see the Appendix for the derivation). However, there is also a limit to this efficiency ratio of a next- to previous-level innovation system: the sum of the series remains finite if this ratio does not exceed 20%. The relative TH to DH capacity then equals maximally:

$P_3/P_2 = 3.3$. This is very close to the value for $k_3/k_2 = 3.24$ that we found above as the efficiency ratio when comparing the periods 1979-2006 with 1945-1980 in terms of US patents per million inhabitants.

The increase in capacity in a TH system can be considered as the result of non-linear interactions within the system which cause auto-catalytic self-organization and effectively increase the complexity of the system's network. In the process of derivation of the model we made some simplifying assumptions for technical reasons, such as using average efficiencies of innovation systems. However, relaxation of these assumptions cannot be expected to change the basic (mathematical) properties of the model.

## 4. Future perspectives

The slope of the curve in Figure 1 indicates a further acceleration of the patent generation speed since 2006 which suggests the change in the knowledge-generating paradigm. Such a change also implies structural change of the corresponding innovation system carrying the knowledge generation. Since the number of patents per million inhabitants is rapidly expanding during the last few years, the potential paradigm extension may indicate the introduction of additional dimensions into the model.

In the innovation-studies literature, various attempts have been made to extend the TH model to a Quadruple Helix (QH) [55, 56, 57, 58, 59] and Quintuple Helix models [60]. Leydesdorff [61] argued that even *N*-tuple helices can be modeled, but suggested to extend the model stepwise in accordance to the explanatory power needed. He noted that the higher dimensions would require separate specification and operationalization in terms of potentially relevant data, and should be empirically actualized in terms of the current economic conditions. Ivanova [62] discussed a possible extension using the media as a fourth dimension of innovation systems as depicted in Figure 13.

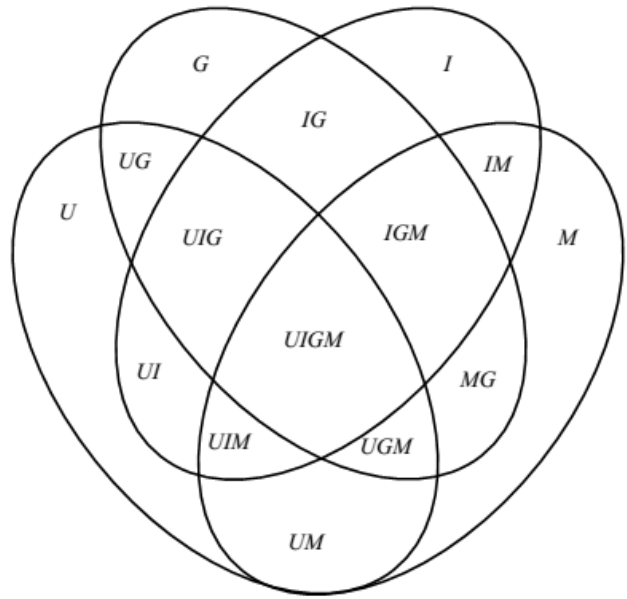

**Fig.13:** Quadruple Helix model of university (*U*), industry (*I*), government (*G*), and media (*M*) relations. (Source: [63, 64]).

The topicality of the above QH model is supported by empirical evidence of the historical presence of the corresponding bi-lateral and three-lateral interactions. Whereas the areas *UI*, *IG*, *UG*, *UIG* refer to the original TH model (Figure 8 above), *IM* refers to commercial advertising by manufacturers, *UM* to public provision of information by science (for example, with the objective to obtain funding), and *MG* to the relation between Media and the State. The intersection *IGM* can be considered as another TH in which the University as actor is replaced by its functional equivalent—an actor Media (M) [23]. *UIM* is widely used in modern business practice when University-Industry relations also include informational (Media) components. In *UGM* relations actor Media replaces as another selection environment the actor Industry. This can be realized in government-funded cultural organizations, etc. In summary, one can state that the post-industrial economy calls for an additional component of information flows to be included in innovation system [12].

More complex arrangements can be constructed bottom-up when all components are historically in place. Similar to the TH case, a QH phase space can abstractly be represented in a four-dimensional orthogonal coordinate system, spanned along the axes *U*, *I*, *G*, *M*. This phase space can be depicted as a three-dimensional spherical surface which takes the form of concave spherical tetrahedron with dihedral angles $\pi/2$. For the purpose of illustration, this system can be considered as a normal tetrahedron. QH, like TH system, possesses non-Abelian (non-commutative) rotational symmetry, associated with a group of rotations in 4D space *O(4)*. The non-commutative rotational symmetry means that the order of two successive changes cannot be changed without changing the result of the operation. From the perspective of the mathematical model [25] this condition ensures the non-linear system dynamics.

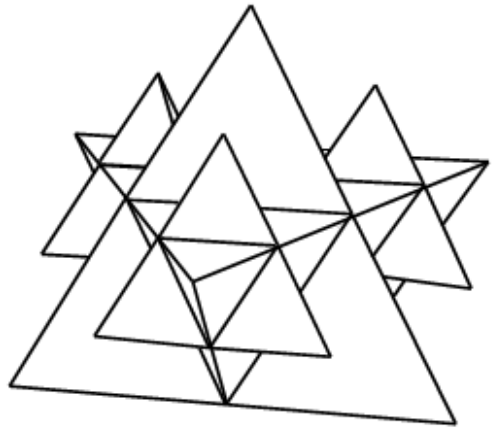

**Fig.14:** Third pre-fractal of the 3D Koch snowflake

In such a configuration, a self-organization of the system can be expected that appears in the form of newly generated networks at different scale levels. Schematically, the QH phase space can then be depicted as a 3D Koch snowflake which is constructed in a manner, similar to that used in the construction of a 2D Koch snowflake, by starting with a regular tetrahedron and iteratively building smaller-sized tetrahedrons at each of surfaces. The 3D Koch snowflake is the limit approached when the above steps are infinitely recursive. The second iteration of an emerging fractal manifold is shown in Figure 14.

If one assumes that in the iterations the average volume of next-level tetrahedron equals $1/\gamma$ of the tetrahedron volume at the previous level, then the level of next to previous system efficiency ratio is preserved. Like in the TH case, one can derive an average capacity of next to previous for the quadruple helix system: this ratio should not exceed 1/7 (approximately 14%; see the Appendix for the derivation).

One interesting consequence of the assumption of fractal structures in the model is that the time cycles of the knowledge-generating paradigm can be analyzed in terms of Fibonacci numbers. This regularity can provide us with a prediction of expected paradigm changes.

Fibonacci numbers are a sequence where each successive number is the sum of the two previous ones: $C_i = C_{i-1} + C_{i-2}$:

$$0, 1, 1, 2, 3, 5, 8, 13, 21, 34, 55, 89, 144 \ldots \qquad (2)$$

One can make the ratio of the current Fibonacci number to the next one: $F_1 = C_i / C_{i+1}$, to one two positions down: $F_2 = C_i / C_{i+2}$, to one three positions down $F_3 = C_i / C_{i+3}$, and so on. The limits of these ratios when *i* increases infinitely are often used in stock and FOREX markets analysis to compare market price movements to one another [65]. The first few such ratio limits are:

$$0.618, \mathit{0.382, 0.236,} 0.145, 0.09 \ldots \qquad (3)$$

Time intervals of the first three periods (cycles), distinguished from Figure 1, and described as different knowledge-generating paradigms correspondingly equal 105, 35, and 25 years. Forming the ratio of the second to first time interval (in Figures 2 and 3): 35/105= 0.33, and of the third to the second cycle time interval (in Figures 3 and 4): 25/105 = 0.238, one can observe that these empirical values correspond closely to the second and third terms of the Fibonacci ratio values sequence (Eq. 3).

Assuming that the same mechanisms continue to operate, the next change in the knowledge-generating paradigm can be expected at 105*0.145 = 15 years from 2006 (that is, in 2021) and at 105*0.09 = 9.5 years from 2021 (that is, in 2030). Changes in the knowledge-generating paradigm imply acceleration in the reflexive changes in the corresponding carriers (that is, innovation systems) given the global selection pressure of a new knowledge-generating paradigm for institutional adjustments [50].

## 5. Conclusions and discussion

The results obtained in the previous sections can be summarized as follows:

1. The process of knowledge generation shows a cyclic nature. However, the cycles shown in terms of numbers of patents per million inhabitants only partly correspond to economic cycles. The differences can be attributed to a different dynamics in economic and knowledge-generating cycles;
2. The knowledge-generating efficiency in successive cycles accelerates;
3. Using a model, the increased knowledge-generating capacity could be explained in terms of the increased complexity of the corresponding innovation systems. This explanation is based on the specific topology of the innovation systems in question. When all relations are historically in place, the increased complexity induces the system's self-organization of a next dimension. This topological extension boosts the knowledge-generating efficiency;
4. The duration of each successive knowledge-generating system is less than the duration (sustainability?) of the previous ones while the capacity accordingly increases; unlike business cycles, knowledge-generating systems tend to exhibit ever shorter cycles;
5. The corresponding time cycles of knowledge-generating paradigms were approximated in terms of Fibonacci numbers and this perspective also enabled us to forecast expected dates of future changes at the system's level.

Some questions need additional investigation. Whether the approximation of the average weight function value $\bar{f}$ remains the same across the different knowledge-generating systems and whether the ratio of the next to the previous ratio of capacity of innovations system does not empirically exceed 20% in the TH case or 14% in the QH case? The most direct test of this implication of the model would come from detailed case studies of patent output at different scale levels and produced by different types (respectively TH and QH) of innovation systems.

Although the model suggests an approach to evaluating the efficiency of innovation systems that can be standardized, the results of innovation activity can nevertheless be expected to differ across countries. Institutional adjustment is historically contingent. However, the question arises inevitably which innovation systems are most efficient and why? For example, when comparing regions, one can ask what are the reasons of differences in the obtained indicator values (in terms of numbers of innovations, inventions, or patents)? A plausible answer would be that the quality of the relations among actors is different, but this would require historical specification. We used USPTO data because the US patent system is often considered as the most competitive one [66].

## 6. Policy implications

Taking a closer look at the foundations of sustainable economic development one can discern a multi-layered structure. Economy growth, as commonly acknowledged [67, 68], depends on the production of knowledge, new technologies, and innovations. But the mechanisms responsible for knowledge production are also the subject of evolution. One can expect that the complexity of newly emerging knowledge-generating systems is successively and step-wise increased, when compared to their predecessors. The time intervals between structural changes in the knowledge-generating systems can be expected to decrease inversely. The shift in the economic dynamics also entails changes in the knowledge-generating paradigm (and vice versa).

This dynamic of a knowledge-generating system that self-organizes its social impact, was not present in an industrial economy. The transition from industrial to post-industrial stages made it manifest. Whereas an industrial economy supported technological trajectories which were relatively stable over time (e.g., railway systems), a complex innovation system, as it is inherent to a knowledge-based economy, implies increasingly diversified technologies, with ever shorter life-cycles. If the life-cycles of innovation systems have a tendency to become shorter, not only adherence to particular technologies, but also adherence to a particular innovation system may lead to a loss in competitive advantage. Therefore, innovation policies under this accelerating regime of changes in economic and knowledge-generating paradigms should focus not so much on manufacturing technologies, but on developing new organizational formats of human interactions that can facilitate the institutional adjustments that can be envisaged as functional.

**References**

[1] R. Nelson, S. Winter, An Evolutionary Theory of Economic Change, Belknap Press of Harvard University Press, Cambridge, MA,1982.

[2] E. S. Andersen, Evolutionary Economics: Post-Schumpeterian Contributions. Pinter, London, 1994.

[3] J. Schumpeter, Business Cycles: A Theoretical, Historical and Statistical Analysis of Capitalist Process, McGraw-Hill, New York, [1939]1964.

[4] P. Romer, Increasing returns and long-run growth, J. Polit. Econ. 94(5) (1986) 1002-1037.

[5] P. Romer, Endogenous technological change, J. Polit. Econ. 98(5) (1990) S71-102.

[6] C. Freeman, The long wave debate, in: T. Vasko (Ed.), Technical Innovations, Diffusion and long Cycles of Economic Development, Springer, Berlin, 1987.

[7] B.-Å. Lundvall, Innovation as an interactive process: from user-producer interaction to the national system of innovation, in: G. Dosi, C. Freeman, R. Nelson, G. Silverberg & L. Soete (Eds.), Technical Change and Economic Theory, Pinter, London, 1988

[8] B.-Å. Lundvall, (Ed.), National Systems of Innovation, Pinter, London, 1992.

[9] R. R. Nelson, (Ed.), National Innovation Systems: A comparative analysis, Oxford University Press, New York, 1993.

[10] M. Porter, The Competitive Advantage of Nations, Free Press, New York, NY, 1990.

[11] M. Porter, On Competition, Harvard Business School, Boston, MA, 1998.

[12] M. Gibbons, C. Limoges, H. Nowotny, S. Schwartzman, P. Scott, M. Trow, The new production of knowledge, The Dynamics of Science and Research in Contemporary Societies, SAGE, London, 1994.

[13] H. Etzkowitz, L. Leydesdorff, The Triple Helix—University-Industry-Government Relations: a laboratory for knowledge-based economic development. EASST Review 14 (1995) 14–19.

[14] H. Etzkowitz, L. Leydesdorff, The dynamics of innovation: from National Systems and "Mode 2" to a Triple Helix of university-industry-government relations. Res. Policy, 29 (2000) 109-123.

[15] H. J. Braczyk, P. Cooke, M. Heidenreich, (Eds.), Regional Innovation Systems, University College London Press, London/ Bristol PA, 1998.

[16] P. Cooke, Knowledge Economies, Routledge, London, 2002.

[17] S. Breschi, F. Malerba, Sectoral innovation systems, in: C. Edquist (Ed.) Systems of Innovation: Technologies, Institutions and Organizations, Pinter, London, 1997.

[18] F. Malerba, Sectoral systems of innovation: a framework for linking innovation to the knowledge base, structure and dynamics of sectors, Econ. Innov. New Technol. 14(1-2) (2005) 63-82.

[19] B. Carlsson, R. Stankiewitz, On the nature, function and composition of technological systems, J. Evol. Econ. 1(1991) 93–118.

[20] B. Carlsson, Internationalization of Innovation Systems: A Survey of the Literature. Res. Policy 35(1) (2006) 56-67.

[21] O. Granstrand, Corporate innovation systems, A Comparative Study of Multi-Technology Corporations in Japan, Sweden and the USA. Chalmers University of Technology, 2000.

[22] B. Lengyel & L. Leydesdorff, Regional innovation systems in Hungary: The failing synergy at the national level, Reg. Stud. 45(5) (2011) 677-693.

[23] H. Etzkowitz and L. Raiken, Artist social movements of the 1960's and 70's: from protest to institution formation. Eric.ed.gov ERIC No. ED186326 ,1980.

[24] G. Dosi, Technological paradigms and technological trajectories: a suggested interpretation of the determinants and directions of technical change, Res. Policy 11 (1982) 147–162.

[25] I. Ivanova, L. Leydesdorff, Rotational symmetry and the transformation of innovation systems in a Triple Helix of university-industry-government relations. Technol. Forecast. Soc. Change, 2014 (in press) http://dx.doi.org/10.1016/j.techfore.2013.08.022

[26] G. Simmel, Individual and Society, in: K. H. Wolf (Ed.), The Sociology of George Simmel, New York: Free Press, 1950.

[27] Y. Cai, Factors Affecting the Efficiency of the BRICS's National Innovation Systems: A Comparative Study based on DEA and Panel Data Analysis, (2011), http://www.economics-ejournal.org/economics/discussionpapers/2011-52 Accessed on April, 7, 2014.

[28] H. Hollanders, A. van Cruysen, Rethinking the European Innovation Scoreboard: A New Methodology for 2008-2010, Report to Pro-INNO, European Commission DG Enterprise and Industry, 2008.

[29] H. Romijn, M. Albaladejo, Determinants of Innovation Capability in Small Electronics and Software Firms in Southeast England, Res. Policy 31 (2002) 1053-67.

[30] M. Porter, S. Stern, National Innovative Capacity. In: Porter, M. E., Sachs, J. D., Cornelius, P. K., McArthur, J. W., Schwab, K. eds. The Global Competitiveness Report 2001–2002, Oxford University Press, New York, 2002, pp. 102-118.

[31] M. J. Farrell, The Measurement of Productive Efficiency, J. Roy. Stat. Soc. 120 (1957) 253-282.

[32] M. Fritsch, V. Slavtchev, How does industry specialization affect the efficiency of regional innovation systems? Ann. Regional Sci. 45 (1) (2010) 87-108.

[33] J. Schmookler, Economic Sources of Inventive Activity, J. Econ. Hist. 22 (1962) 1-20.

[34] Z. Griliches, Issues in Assessing the Contribution of Research and Development to Productivity Growth, Bell J. Econ. 10 (1979): 92-116.

[35] A. B. Jaffe, Real Effects of Academic Research, Am. Econ. Rev. 79 (1989) 957-970.

[36] R. D. Shelton, L. Leydesdorff, Publish or Patent: Bibliometric evidence for empirical trade-offs in national funding strategies, J. Am.Soc. Inf. Sci. Technol. 63(3) (2012). 498-511.

[37] L. Leydesdorff, M. Fritsch, Measuring the knowledge base of regional innovation systems in Germany in terms of a Triple Helix dynamics, Res. Policy 35 (2006) 1538-1553.

[38] K. Tominomori, Self-Organization Theory and Its Applicability to "Economic System" Econ. J. of Hokkaido Univ. 31 (2002) 41-62.

[39] R. M. Solow, Technical progress and the aggregate production function. Rev.Econ. Stat. 39 (1957) 312-20.

[40] D.T. Coe, E. Helpman, A.W. Hoffmaister. International R&D spillovers and institutions. Eur. Econ. Rev. 53 (7) (2009) 723-41.

[41] M. Fritsch, V. Slavtchev, Measuring the Efficiency of Regional Innovation Systems An Empirical Assesment, (2006), http://tu-freiberg.de/sites/default/files/media/fakultaet-6-3307/fileadmin/Arbeitspapiere/2006/fritsch_8__2006.pdf Accessed on April, 7, 2014

[42] A. Korotaev, Ju. Zinkina, Ju. Bogevolnov, Kondratieff waves in global invention activity (1900-2008), Technol. Forecast. Soc. Change. 78, (2011) 1280-1284.25

[43] C. Freeman, As Time Goes By, Oxford, Oxford University Press, 2001.

[44] R. Aires, Technological transformations and long waves, (1989) http://webarchive.iiasa.ac.at/Admin/PUB/Documents/RR-89-001.pdf Accessed on April, 7, 2014

[45] Y. Jung, N. Vonortas, Techno-Economic Paradigm Shift and Contemporary Justification of STI Policy in Korea and the United States, (2002), http://www.gwu.edu/~cistp/assets/docs/research/papers/Jang%20-%20Paradigm%20%28STEPI%29%202002.with.Vonortas.pdf Accessed on April, 7, 2014

[46] OECD Science, Technology and Industry Scoreboard 2000. Paris, Organization for Economic Cooperation and Development, OECD (2000).

[47] Globalization and Competitiveness: Relevant Indicators. Paris, Directorate for Science, Technology and Industry: 60, OECD (1996).

[48] The New Economy: Beyond the Hype Final Report on the OECD Growth Project: Executive Summary. Paris, Organization for Economic Cooperation and Development: 26, OECD (2001).

[49] H. Etzkowitz & L. Leydesdorff, The endless transition: a "triple helix" of university–industry–government relations. Minerva, 36, (1998) 203–208.

[50] C. Freeman, C. Perez, Structural crisis of adjustment, business cycles and investment behavior, in: G. Dosi, et al., (Eds.), Technical Change and Economic Theory, Pinter, London, 1988.

[51] L. Leydesdorff, I. Ivanova, Mutual Redundancies in Inter-human Communication Systems: Steps Towards a Calculus of Processing Meaning, J. Assoc. Inf. Sci. Technol. 65(2) (2014) 386-399.

[52] L. Leydesdorff & G. Zawdie, The Triple Helix Perspective of Innovation Systems, Technol. Anal. Strateg. Management 22(7) (2010) 789-804

[53] A. de Rosa Pires & E. de Castro, Can a Strategic Project for a University Be Strategic to Regional Development? Sci. Publ. Policy 24(1) (1997) 15-20.

[54] B. Mandelbrot, W. Freeman, The Fractal Geometry of Nature, W.H. Freeman, San Francisco, 1983.

[55] Z. Baber, Globalization and scientific research: the emerging triple helix of state-industry-university relations in Japan and Singapore. B. Sci. Technol. Soc. 21, (2001) 401–408.

[56] J. Bunders, , J. Broerse, M. Zweekhorst, The triple helix enriched with the user perspective: A view from Bangladesh, J. Technol. Transfer, 24(2) (1999) 235- 246.

[57] E. G. Carayannis & D. F. J. Campbell, 'Mode 3'and'Quadruple Helix': toward a 21st century fractal innovation ecosystem. Int. J. Technol. Manage. 46(3) (2009) 201–234.

[58] L. Leydesdorff, Yuan Sun, National and International Dimensions of the Triple Helix in Japan: University-Industry-Government *versus* International Co-Authorship relations, J. Am. Soc. Inf. Sci. Technol. 60(4) (2009) 778-788.

[59] M. Mehta, Regulating biotechnology and nanotechnology in Canada: a post-normal science approach for inclusion of the fourth helix, in: Z. Baber and H. Klondker (Eds.), The Triple Helix, Albany: State University of New York Press, 2003.

[60] E. G. Carayannis, D. F. J. Campbell, Triple Helix, Quadruple Helix and Quintuple Helix and how do knowledge, innovation, and environment relate to each other? Int. J. Soc. Ecol. Sustain. Devel. 1(1) (2010) 41–69.

[61] L. Leydesdorff, The Triple Helix, Quadruple Helix…, and an N-tuple of Helices: explanatory models for analyzing the knowledge-based economy, J. Knowl. Econ. 3(1) (2012).  25–35.

[62] I. Ivanova, Quadruple Helix System and Symmetry: a Step Towards Helix Innovation System Classification, J. Knowl. Econ. (2014) DOI 10.1007/s13132-014-0201-z

[63] R. Rousseau, Venn, Caroll, Karnaugh en Edwards. Wiskunde & Onderwijs 24(95) (1998) 233–241.

[64] Durrett, R. (s.d.). Mathematics behind numb3rs. Season 4, Episode 12: Power. Available at http://www.math.cornell.edu/~numb3rs/lipa/ power.html

[65] R. W. Colby, The Encyclopedia of Technical Market Indicators, McGraw-Hill Publishing, 2003.

[66] F. Narin, K. S. Hamilton, D. Olivastro, The increasing link between U.S. technology and public science. Res. Policy, 26(3) (1997) 317-330.

[67] N. Sakurai, G, Papaconstantinou, E. Ioannidis, The Impact of R&D and Technology Diffusion on Productivity Growth: Evidence from Ten OECD Countries in the 1970s and 1980s, STI Working Papers 1996/2, OECD, Paris, 1996.

[68] P. Romer, The Origins of Endogenous Growth, J. Econ. Perspect. 8(1) (1994) 3-22.

[69] A. Kolpakov, A. Mednykh, M. Pashkevich, Volume formula for a $\mathbb{Z}_2$-symmetric spherical tetrahedron through its edge lengths. Ark. Mat. 51(1) (2013) 99-123.

# Appendices

## A. The 2D Koch snowflake

Let the surface area of each next triangle in the iteration triangle equals $1/\beta$ of the previous one triangle area. The surface surplus to the initial triangle produced by second, third, and forth iterations shown in Fig. 4 correspondingly equal:

$$S_1 = \frac{3}{\beta}S$$

$$S_2 = S_1 \cdot 5S\,{}^1/_\beta \qquad (A.2)$$

$$S_3 = S_1 \cdot 5 \cdot 5\left({}^1/_\beta\right)^2$$

One can iteratively continue and get

$$S_n = S_1\left({}^5/_\beta\right)^{n-1} \qquad (A.3)$$

The supplement area

$$\Sigma = S_1 + S_2 + \cdots + S_n + \cdots \qquad (A.4)$$

can be calculated according the formula for geometric progression with common ratio $r = {}^5/_\beta$ and scale factor $b = S_1$. If $\beta = 6.3$

$$\Sigma = \lim_{n\to\infty} b\frac{1-r^n}{1-r} = S_1 \frac{1}{1-{}^5/_\beta} = 2.3S \qquad (A.5)$$

This means that the area of Koch snowflake exceeds the area of initial triangle by a factor 3.5. In order the sum of the series converges $\beta$ should be less than 5.

## B. The 3D Koch snowflake

The 3D Koch snowflake is constructed on the base of regular spherical tetrahedron of the unit sphere in which all four faces are equilateral spherical triangles. The volume of $\mathbb{Z}_2$-symmetric spherical tetrahedron with dihedral angles $\pi/2$ is [69]

$$V = \pi^2/8 \tag{B.1}$$

After each iteration a new regular tetrahedron is added on each face of the previous iteration. One can get:

$$V_1 = 4V \, 1/\gamma$$

$$V_2 = 4 \cdot 7V(1/\gamma)^2 \tag{B.2}$$

$$V_3 = 4 \cdot 7 \cdot 7V(1/\gamma)^3$$

$$\ldots$$

$$V_n = V_1 \cdot (7/\gamma)^{n-1}$$

The total volume increment to the original snowflake after $n$ iterations

$$\Omega_n = V_1 + V_2 + \cdots + V_n$$

Collapsing the geometric sum gives:

$$\Omega = 4V \tag{B.3}$$

the volume of Koch snowflake exceeds the volume of initial tetrahedron by a factor $1 + 4/(\gamma - 1)$. In order the sum of the series converges $\gamma$ should be less than 7.